\documentclass[aps,pra,twocolumn,superscriptaddress,showpacs]{revtex4}
\usepackage{amssymb}
\usepackage{amsmath}
\usepackage{graphicx}
\usepackage{color}

\definecolor{red}{rgb}{1,0,0}
\input{tcilatex}

\begin{document}

\title{He-LiF surface potential from fast atom diffraction under grazing
incidence.}
\author{A. Sch{\"u}ller}
\affiliation{Institut f\"ur Physik, Humboldt Universit\"at zu Berlin,\\
Newtonstrasse 15, D-12489 Berlin-Adlershof, Germany.}
\author{H. Winter}
\affiliation{Institut f\"ur Physik, Humboldt Universit\"at zu Berlin,\\
Newtonstrasse 15, D-12489 Berlin-Adlershof, Germany.}
\author{M.S. Gravielle\thanks{%
Author to whom correspondence should be addressed.\newline
Electronic address: msilvia@iafe.uba.ar}}
\affiliation{Instituto de Astronom\'{\i}a y F\'{\i}sica del Espacio. Consejo Nacional de
Investigaciones Cient\'{\i}ficas y T\'{e}cnicas. Casilla de Correo 67,
Sucursal 28, (C1428EGA) Buenos Aires, Argentina.}
\affiliation{Departamento de F\'{\i}sica. Facultad de Ciencias Exactas y Naturales.
Universidad de Buenos Aires. }
\author{J.M. Pruneda}
\affiliation{Centre d'Investigaci\'{o} en Nanoci\`{e}ncia i Nanotecnolog\'{\i}a-CIN2
(CSIC-ICN), Campus UAB 08193 Bellaterra, Spain}
\author{J.E. Miraglia}
\affiliation{Instituto de Astronom\'{\i}a y F\'{\i}sica del Espacio. Consejo Nacional de
Investigaciones Cient\'{\i}ficas y T\'{e}cnicas. Casilla de Correo 67,
Sucursal 28, (C1428EGA) Buenos Aires, Argentina.}
\affiliation{Departamento de F\'{\i}sica. Facultad de Ciencias Exactas y Naturales.
Universidad de Buenos Aires. }
\date{\today }

\begin{abstract}
Diffraction patterns produced by grazing scattering of fast atoms from
insulator surfaces are used to examine the atom-surface interaction. The
method is applied to He atoms colliding with a LiF(001) surface along axial
crystallographic channels. The projectile-surface potential is obtained from
an accurate DFT calculation, which includes polarization and surface
relaxation. For the description of the collision process we employ the
surface eikonal approximation, which takes into account quantum interference
between different projectile paths. The dependence of projectile spectra on
the parallel and perpendicular incident energies is experimentally and
theoretically analyzed, determining the range of applicability of the
proposed model.
\end{abstract}

\maketitle

\section{Introduction}

Diffraction of thermal atoms from crystal surfaces has been extensively
studied over the years \cite%
{Cabrera70,Boato73,Wolken1973,Boato1976a,Garcia1976b,Garcia1977b,Hubbard83,Celli85,Ekinci2004,Jardine2004,Riley2007}%
, becoming a common tool for surface analysis. Recently new experiments \cite%
{Schuller07,Rousseau07,Schuller08,Schuller2009} \ have shown interference
effects also for grazing scattering of fast atoms from surfaces, where
classical mechanics was supposed to be adequate. This unexpected diffraction
phenomenon was found to be very sensitive to the description of the
projectile-surface interaction \cite%
{Schuller08,Schuller2009,Gravielle08,Aigner08,Schuller2009PRB} which opens
the way for a method to probe surface potentials with high accuracy.

The aim of this work is to find out to what extent surface potentials
derived from state-of-art \textit{ab-initio} methods are capable of
reproducing experimental diffraction patterns for grazing scattering of
swift He atoms from a LiF(001) surface. The He-LiF surface interaction is
here derived by using the \emph{Siesta }\cite{siesta} implementation of the
density-functional theory (DFT), which is a self-consistent method for
performing first-principles calculations on systems with a large number of
atoms. This DFT method has been successfully used to study a variety of
nanoscale problems \cite{siesta1}. In order to describe the interference
process, we employ a distorted-wave model - the surface eikonal
approximation \cite{Gravielle08} - using the eikonal wave function to
represent the elastic collision with the surface, while the motion of the
fast projectile is described classically by considering axially channeled
trajectories for different initial conditions. The surface eikonal
approximation is valid for small de Broglie wavelengths of incident atoms,
as considered here, which are several orders of magnitude smaller than the
interatomic distances in the crystal. This method was shown to provide an
adequate description of the interference effects for atoms colliding with
insulator surfaces under axial surface channeling \cite{Gravielle09}.

Eikonal projectile angular distributions derived from using the DFT surface
potential are compared with the experiment for different energies of
incident projectiles. From this comparison we deduce the validity range of
the potential model, which involves polarization and surface rumpling. The
paper is organized as follows. The experimental \ method and the theoretical
formalism are summarized in Sec. II and III, respectively. Results are
presented and discussed in Sec. IV, and in Sec. V we outline our
conclusions. Atomic units (a.u.) are used unless otherwise stated.\medskip

\section{Experimental method}

In our experiments we have scattered neutral $^3$He and $^4$He atoms with
kinetic energies $E_{i}$ ranging from 0.3 keV to 25 keV from a clean and
flat LiF(001) surface at room temperature under grazing angles of incidence $%
0.4< \theta_i < 1.5$ deg. Fast He$^+$ ion beams were produced in a 10 GHz
electron cyclotron resonance (ECR) ion source (Nanogan Pantechnique, Caen,
France). The neutralization of the He$^+$ ions was achieved via charge
transfer in a gas cell mounted in the beam line of the accelerator operating
with He gas and subsequent deflection of remaining charged fraction by an
electric field. A base pressure of some 10$^{-11}$ mbar was achieved in our
UHV chamber by a turbomolecular pump in series with a titanium sublimation
pump, where the pressure gradient with respect to the beam line of the
accelerator was maintained by two differential pumping stages. Pairs of
slits at both ends of these stages were used for the collimation of the
incident beam to a divergence of $<0.03^\circ$. This high collimation is
necessary for diffraction in order to maintain the degree of coherence in
the scattering process from LiF(001) .

The LiF(001) surface was prepared by cycles of grazing sputtering with 25
keV Ar$^{+}$ ions at $250^{\circ }$C where the ionic conductivity of LiF is
sufficiently enhanced in order to avoided macroscopic charging up and
subsequent annealing to temperatures of about $350^{\circ }$C. The
scattering experiments were performed in the regime of axial surface
channeling, i.e the azimuthal setting of the surface plane was chosen so
that the direction of the incident beam was parallel with atomic strings
along low indexed directions in the surface plane.

2D angular distributions of scattered projectiles were recorded by means of
a commercially available position-sensitive multi-channelplate detector
(MCP) with a delay-line anode (DLD40, Roentdek Handels GmbH) located 66 cm
behind the target. This provides a simple and very efficient procedure for
recording data where complete diffraction patterns as shown below can be
recorded in a time of about minutes. Since only about 10$^4$ He atoms per
second hit the target surface, fast atom diffraction is non-destructive and
can be applied in studies on insulator surfaces (neutral projectiles) \cite%
{Schuller07,Rousseau07}, as well as adsorption phenomena at metal surfaces 
\cite{Schuller2009,Schuller2009PRB}.

Since aside from the absolute angular positions of diffraction spots, their
relative intensities are important here, one has to carefully correct the
recorded diffraction patterns with respect to inhomogeneities in the
detection efficiency across the sensitive area of the MCP. This correction
is performed by means of a wobbling technique where the projectile beam is
scanned across the MCP active area with two orthogonal oriented electric
fields using frequencies in the kHz domain.

\section{Theoretical model}

When an atomic projectile ($P$) impinges under grazing incidence on a
crystal surface ($S$) the T- matrix element associated with the elastic
scattering can be defined in terms of the scattering state of the
projectile, $\Psi _{i}^{+}$, as 
\begin{equation}
T_{if}=\int d\vec{R}_{P}\ \Phi _{f}^{^{\ast }}(\vec{R}_{P})\ V_{SP}(\vec{R}%
_{P})\Psi _{i}^{+}(\vec{R}_{P}),  \label{Tif}
\end{equation}%
where $V_{SP}$ is the surface-projectile interaction, $\vec{R}_{P}\ $denotes
the position of the center of mass of the incident atom, and $\Phi _{j}(\vec{%
R}_{P})=(2\pi )^{-3/2}\exp (i\vec{K}_{j}\cdot \vec{R}_{P})$, with $j=i(f)$,
being the initial (final) unperturbed wave function and $\vec{K}_{i(f)}$ the
initial (final) projectile momentum. Taking into account that in the range
of impact energies the de Broglie wavelength of the incident projectile, $%
\lambda =2\pi /K_{i}$, is sufficiently short compared to the characteristic
distance of the surface potential, we approximate the scattering state $\Psi
_{i}^{+}$ by means of the eikonal-Maslov wave function \cite{Joachain} as
follows 
\begin{equation}
\Psi _{i}^{+}(\mathcal{\vec{R}}_{P})\simeq \chi _{i}^{^{(eik)+}}(\mathcal{%
\vec{R}}_{P})=\Phi _{i}(\mathcal{\vec{R}}_{P})\exp (-i\eta (\mathcal{\vec{R}}%
_{P})),  \label{feik}
\end{equation}%
where 
\begin{equation}
\eta (\mathcal{\vec{R}}_{P}(t))=\int\limits_{-\infty }^{t}dt^{\prime }\
V_{SP}(\mathcal{\vec{R}}_{P}(t^{\prime }))+\phi _{M}  \label{fasen}
\end{equation}%
is the eikonal-Maslov phase, which depends on the classical position of the
incident atom $\mathcal{\vec{R}}_{P}$ at a given time $t$. This phase
includes the Maslov correction term $\phi _{M}=\nu \pi /2$ that takes into
account the phase change suffered by the projectile at turning points, with $%
\nu $ the Maslov index defined as in Ref.\cite{Avrin94}. By inserting Eq. (%
\ref{feik}) in Eq. (\ref{Tif}), after some algebra the eikonal transition
matrix reads \cite{Gravielle08} 
\begin{equation}
T_{if}^{(eik)}=\int d\vec{R}_{os}\ a_{if}(\vec{R}_{os}),  \label{Teikn}
\end{equation}%
where $\vec{R}_{os}$ determines the initial position of the projectile on
the surface plane and%
\begin{eqnarray}
a_{if}(\vec{R}_{os}) &=&\frac{1}{(2\pi )^{3}}\ \int\limits_{-\infty
}^{+\infty }dt\ \left\vert v_{z}(\mathcal{\vec{R}}_{P})\right\vert \times 
\notag \\
&&\exp [-i\vec{Q}.\mathcal{\vec{R}}_{P}-i\eta (\mathcal{\vec{R}}_{P})]\
V_{SP}(\mathcal{\vec{R}}_{P})  \label{aif}
\end{eqnarray}%
is the transition amplitude associated with the classical path $\mathcal{%
\vec{R}}_{P}(\vec{R}_{os},t)$. In Eq. (\ref{aif}) $\vec{Q}=\vec{K}_{f}-\vec{K%
}_{i}$ is the projectile momentum transfer and $v_{z}(\mathcal{\vec{R}}_{P})$
denotes the component of the projectile velocity perpendicular to the
surface plane, with $\hat{z}$ directly along the surface normal, towards the
vacuum region.

The differential probability, per unit of surface area, for elastic
scattering with final momentum $\vec{K}_{f}$ in the direction of the solid
angle $\Omega _{f}\equiv (\theta _{f},\varphi _{f})$ is obtained from Eq. (%
\ref{Teikn}) as $dP/d\Omega _{f}=(2\pi )^{4}m_{P}^{2}\left\vert \tilde{T}%
_{if}^{(eik)}\right\vert ^{2}$, where $\tilde{T}_{if}^{(eik)}$ denotes the
eikonal T-matrix element, normalized per unit area, $m_{P}$ is the
projectile mass, and $\theta _{f}$ and $\varphi _{f}$ are the final polar
and azimuthal angles, respectively, with $\varphi _{f}$ \ measured with
respect to the $\widehat{x}$ axis, along the incidence direction in the
surface plane. Details are given in Refs. \cite{Gravielle08,Gravielle09}.

\subsection{Projectile-surface interaction}

The surface potential was determined by performing first-principles
calculations for the LiF(001) surface. We used the \emph{Siesta }\cite%
{siesta} implementation of density-functional theory (DFT) within the Local
Density Approximation (LDA) \cite{CA} to obtain the effective interaction
potential between a He atom and a slab of 10 atomic planes of LiF. Periodic
boundary conditions were used in the (001) plane, and in order to prevent
the interaction between images of the He atoms a $\sqrt{(}2)\times \sqrt{(}%
2) $ supercell was considered in the (001) plane, giving a total of 112+1
atoms in the simulation box.

We model the core electrons with norm-conserving pseudopotentials of the
Troullier-Martins type \cite{TM}, describing valence electrons with
numerical double-$zeta$ polarized atomic orbitals as the basis set. We
explicitly included semicore states for Li (1s$^{2}$) in the valence band,
and added two extra layers of "ghost orbitals" at the surface to increase
the basis set description of the surface electronic wave-functions.

The structure was optimized until the forces on all atoms were smaller than
0.03 eV/\AA . The slight underestimation for the in-plane lattice constant
obtained in our slab geometry (3.94\AA , experimental 4.02\AA\ \cite%
{Ekinci2004}) is typical of LDA calculations. Different relaxation of the Li
and F atoms at the surface called rumpling is apparent in the first two
atomic layers, with F atoms in the surface plane being slightly pushed out
and Li atoms slightly depressed by a distance $d_{1\text{ }}$measured with
respect to the unreconstructed surface. For the topmost atomic layer we
obtained $d_{1}=0.046$ a.u., while the displacement corresponding to the
second layer is substantially smaller and in opposite direction ($%
d_{2}=-0.010$ a.u.). These values are slightly higher than those reported
from a IV-LEED analysis \cite{Vogt02}.

The surface potential $V(z)$ for a given position $(x,y)$ on the surface
plane was obtained from calculations of the total energy of the system
composed by the LiF slab and the He atom placed a distance $z$ from the last
atomic layer of the slab (taken as the average between Li and F positions).
The standard correction \cite{BSSE} due to BSSE (Basis Set Superposition
Errors) was considered for the computed $V(z)$: 
\begin{equation}
V(z)=E[\text{LiF+He(z)}]-E[\text{He(z)}]^{\text{LiF}}-E[\text{LiF}]^{\text{%
He(z)}},
\end{equation}%
where $E[\text{A}]^{\text{X}}$ denotes the energy of the system A
considering not only the basis orbitals of A, but also those that correspond
to the subsystem X. Once V(z) is known for selected 9 in-plane high-symmetry
positions in a mesh of $z$ points, an interpolation scheme is used to derive 
$V_{SP}(\vec{R})$ at any point in the vacuum region.

\section{Results}

Experimental angular distributions of He$^{0}$ projectiles elastically
scattered from a LiF(001) surface under axial surface channeling conditions
are presented here as a benchmark for the DFT projectile-surface potential.

\FRAME{ftbpFU}{3.5276in}{2.2693in}{0pt}{\Qcb{Two dimensional intensity
distributions, as recorded with a position sensitive detector, for $^{4}$He
atoms scattered from LiF(001) along a $\langle 110\rangle $ direction with
two different projectile energies, $E_{i}$=2.2 keV and 7.5 keV, but with the
same perpendicular energy ($E_{i\perp }$=1.04 eV). Color code: Red=high,
blue=low intensity. Positions of rainbow angles $\Theta _{rb}$ are indicated
by straight lines. Black dots represent theoretical positions of maxima from
the surface eikonal model.}}{\Qlb{Fig2a}}{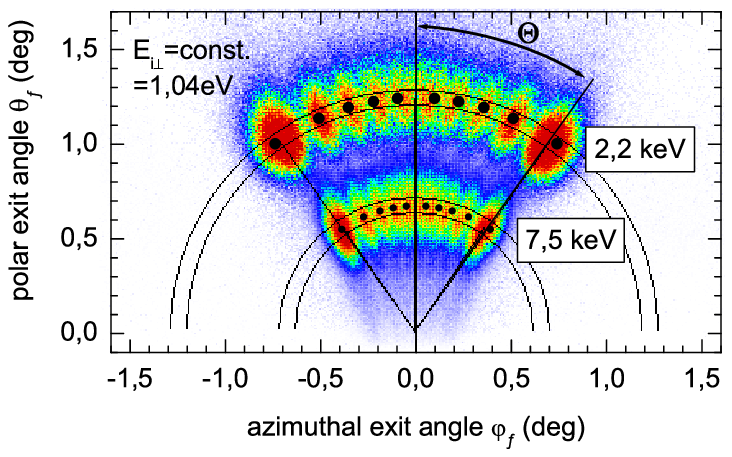}{\special{language
"Scientific Word";type "GRAPHIC";maintain-aspect-ratio TRUE;display
"USEDEF";valid_file "F";width 3.5276in;height 2.2693in;depth
0pt;original-width 3.0978in;original-height 1.983in;cropleft "0";croptop
"1";cropright "1";cropbottom "0";filename 'fig1.eps';file-properties
"XNPEU";}}

First, we analyze the dependence of final projectile distributions on the
incidence energy $E_{i}$, splitted into two terms $E_{i}=E_{i\parallel
}+E_{i\perp }$, where $E_{i\parallel }=E_{i}\cos ^{2}\theta _{i}$ ($%
E_{i\perp }=E\sin ^{2}\theta _{i}$) is associated with the component of the
initial velocity parallel (perpendicular) to the axial channel, with $\theta
_{i}$ the incidence angle measured with respect to the surface plane. In
Fig. \ref{Fig2a} we show diffraction patterns for $^{4}$He atoms impinging
along the $\langle 110\rangle $ direction with two different energies - $%
E_{i}=$ 2.2 keV and 7.5 keV - but with the same perpendicular energy, $%
E_{i\perp }=$1.04 eV. \ In both cases, the distribution for scattered
projectile lies inside an annulus of radius $\theta _{i}$, presenting maxima
symmetrically placed with respect to the incidence direction (i.e. $\varphi
_{f}=0$). The high intensity for the outermost peaks (marked with lines) is
due to rainbow scattering under rainbow angle $\Theta _{rb}$ at maximal
deflection, which can be explained classically. Between the outer rainbow
peaks further peaks show up, which can be explained as quantum mechanical
diffraction effects in close analogy to the origin of supernumerary rainbows 
\cite{Schuller08}. Supernumerary rainbows originate from quantum
interference between projectiles that follow different classical pathways
with the same final momentum. The order $m$ of a supernumerary corresponds
to the multiple of $\lambda $ in path length difference for constructive
interference. The black dots in Fig. \ref{Fig2a} represent the theoretical
positions of the maxima from the eikonal model, which closely agree with
experimental data.

\FRAME{ftbpFU}{204pt}{169.5pt}{0pt}{\Qcb{Projected intensities inside
annuluses in Fig. \protect\ref{Fig2a}, as a function of the deflection angle 
$\Theta $, for $E_{i}$=7.5 keV (gray circles) and 2.2 keV (blue squares) and
corresponding differential probabilities derived from the surface eikonal
approach (black dashed and red full curves). $m$ denotes the order of the
supernumerary rainbow.}}{\Qlb{Fig2b}}{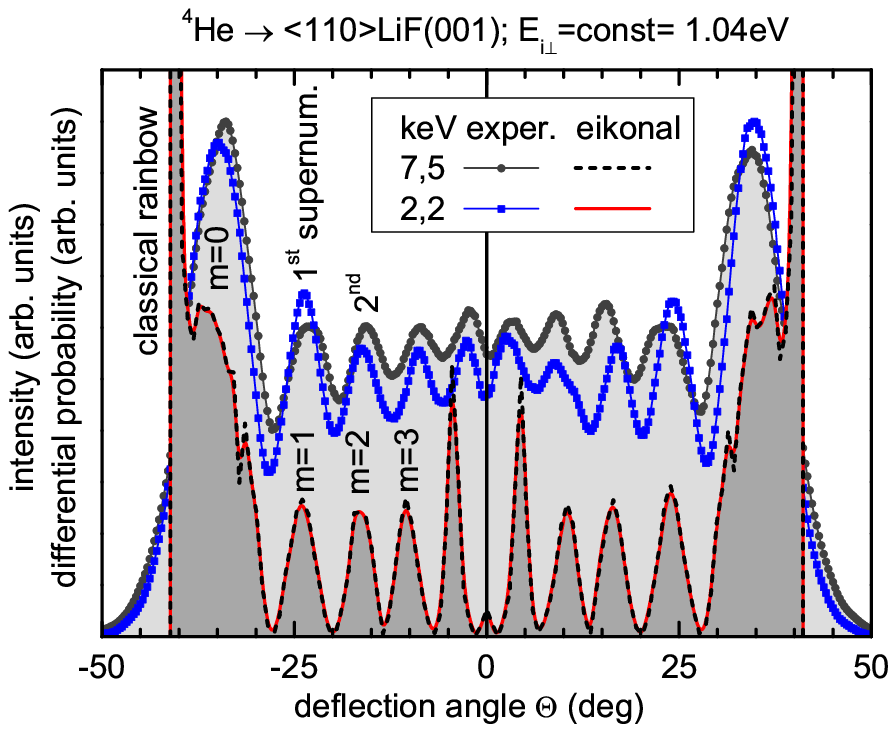}{\special{language
"Scientific Word";type "GRAPHIC";maintain-aspect-ratio TRUE;display
"USEDEF";valid_file "F";width 204pt;height 169.5pt;depth 0pt;original-width
4.1632in;original-height 3.3382in;cropleft "0.0498";croptop
"0.9377";cropright "0.9166";cropbottom "0.0413";filename
'fig2.eps';file-properties "XNPEU";}}

In Fig. \ref{Fig2b} we show the intensity inside the annulus of radius $%
\theta _{i}$ from Fig. \ref{Fig2a} as a function of the deflection angle $%
\Theta $, defined as $\Theta =\arctan (\varphi _{f}\ /\theta _{f})$.
Position and number of the supernumerary maxima are independent of $E_{i}$
at the same $E_{i\perp }$. Similar structures are predicted by the eikonal
model, although in the vicinity of the classical rainbow angle $\Theta _{rb}$
the relative intensity is overestimated. This is because the eikonal model
is a semiclassical method based on classically calculated trajectories,
showing a sharp maximum at the classical rainbow, where intensity increases
sharply for $\Theta \rightarrow \Theta _{rb}$ and is zero for $\Theta
>\Theta _{rb}$. In a more elaborate quantum treatment the classical rainbow
peak will be replaced by a smoother maximum for $m=0$ at $\Theta <\Theta
_{rb}$, with decaying intensity on the dark side of the classical rainbow $%
\Theta >\Theta _{rb}$ \cite{Berry1966,Berry1972}. The maximum $m=0$ is the
\textquotedblleft quantum surface rainbow" \cite{Garibaldi1975}.

\FRAME{ftbpFU}{232.0625pt}{186.25pt}{0pt}{\Qcb{Deflection angles $\Theta $
corresponding to maxima of angular distributions, as a function of the
projectile energy $E_{i}$, for $^{4}$He atoms scattered from LiF(001) along
the direction $\langle 110\rangle $. The perpendicular energy is kept as a
constant ($E_{i\perp }$ =1.04 eV). Circles, experimental data; curves,
quantum rainbow $m=0$ and supernmerary rainbows $m=1$ to 4 derived within
the surface eikonal approximation.}}{\Qlb{Fig3}}{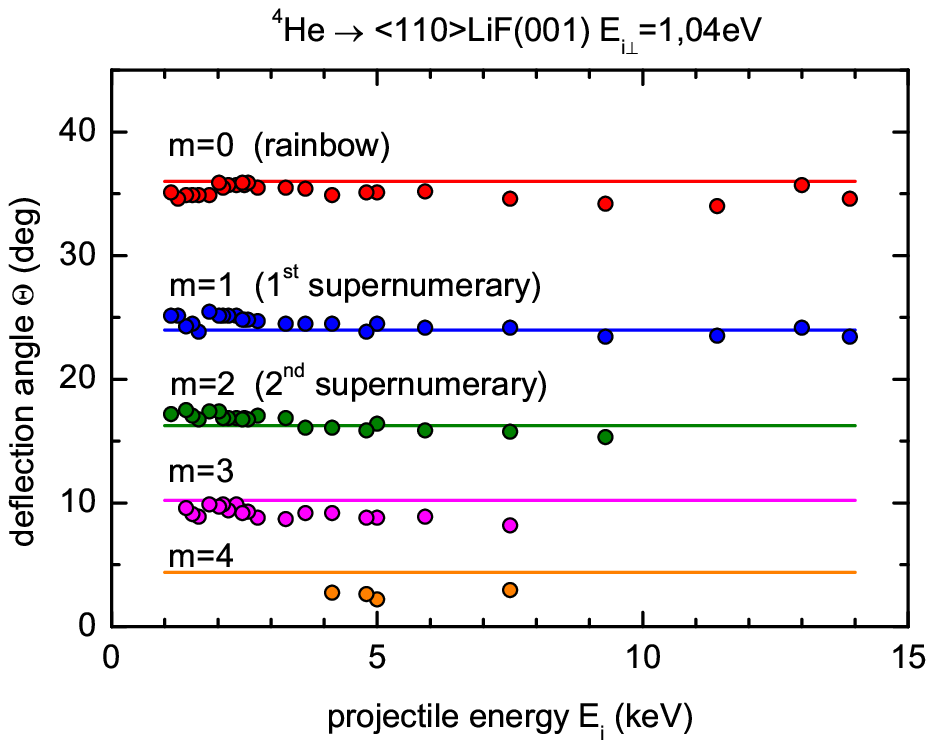}{\special{language
"Scientific Word";type "GRAPHIC";maintain-aspect-ratio TRUE;display
"USEDEF";valid_file "F";width 232.0625pt;height 186.25pt;depth
0pt;original-width 4.3163in;original-height 3.4238in;cropleft
"0.0300";croptop "0.9595";cropright "0.9196";cropbottom "0.0607";filename
'fig3.eps';file-properties "XNPEU";}}

Just like in the experiment, eikonal patterns as a function of the
deflection angle $\Theta $ are independent of $E_{i}$ at the same $E_{i\perp
}$. This becomes more evident in Fig. \ref{Fig3} where deflection angles
corresponding to supernumerary rainbows are plotted as a function of the
total energy $E_{i}$ while keeping $E_{i\perp }$ constant. Even though $%
V_{SP}(\vec{R})$ takes into account the complete corrugation on the surface
plane, without averaging the projectile-surface potential along the
incidence direction, the eikonal projectile distribution is practically
unaffected by the modulation of the potential along the channel. Thus the
differential probability $dP/d\Theta $ is independent of $E_{i\parallel }$
for a given perpendicular energy. On the other hand, the positions of
supernumerary maxima are found to be extremely sensitive to the shape of the
surface potential across the channel, especially for higher orders $m$,
which correspond to small deflection angles \cite%
{Gravielle08,Aigner08,Schuller08,Gravielle09}. From Fig. \ref{Fig3} we show
that the DFT surface potential reproduces fairly well the angular positions
of the supernumeraries.

Different perpendicular energies $E_{i\perp }$ probe a different $z$-range
of $V_{SP}$. To investigate in detail the atom-surface potential across the $%
\langle 110\rangle $ channel, in Fig. \ref{Fig4a} we plot the angular
positions of maxima of the experimental distribution as a function of $%
E_{i\perp }$ ranging from 0.03 to 3 eV. In this case $^{3}$He isotopes are
used as projectiles. For low perpendicular energies, i.e. $E_{i\perp
}\lesssim $ 1 eV, the experimental spectra show maxima at Bragg angles $%
\Theta _{n}$, which fulfill the condition 
\begin{equation}
d\sin \Theta _{n}=n\lambda _{\perp },  \label{bragg}
\end{equation}%
$d$ being the width of the channel, $n$ the diffraction order, and $\lambda
_{\perp }=2\pi /K_{iz}$ the de Broglie wavelength associated with the
perpendicular motion. As discussed in Ref.\cite{Schuller08,Schuller09},
interference patterns for grazing scattering stem from two different
mechanisms. The first one, associated with the supernumerary rainbows, is
produced by the interference of trajectories whose initial positions $\vec{R}%
_{os}$ differ by a distance smaller than $d$, carrying information on the
shape of the interaction potential across the channel. The second one
originates from the interference of trajectories whose initial positions $%
\vec{R}_{os}$ are separated by the spacial lattice periodicity $d$ resulting
in \textquotedblleft Bragg peaks" providing information on the spacing
between surface atoms. Whether a Bragg peak shows intensity or not depends
on the position of the supernumeraries. The Bragg peak of order $n$ which is
closest to the angular position of a supernumerary of order $m$ is intense.
Since the Bragg angles $\Theta _{n}$ decrease with $E_{i\perp }$ while the
angular position of the supernumeraries increases, the order $n$ of the
intense Bragg peak increases successively.

\FRAME{ftbpFU}{247.8125pt}{198.25pt}{0pt}{\Qcb{ Similar to Fig. \protect\ref%
{Fig3}, but as function of the perpendicular energy $E_{i\perp }$, for $^{3}$%
He atoms scattered from LiF(001) along the direction $\langle 110\rangle $.
Full colored curves, quantum rainbow $m=0$ and supernmerary rainbows $m=1$
to 4 derived within the surface eikonal approximation based on the present
DFT potential; dashed curves, positions of supernumerary rainbows in the
hard wall approximation (Eq. (\protect\ref{Bessel})) using the effective
corrugation of the DFT potential across the $\langle 110\rangle $ channel;
grey curves, theoretical positions of maxima from Bragg condition (Eq. (%
\protect\ref{bragg})).}}{\Qlb{Fig4a}}{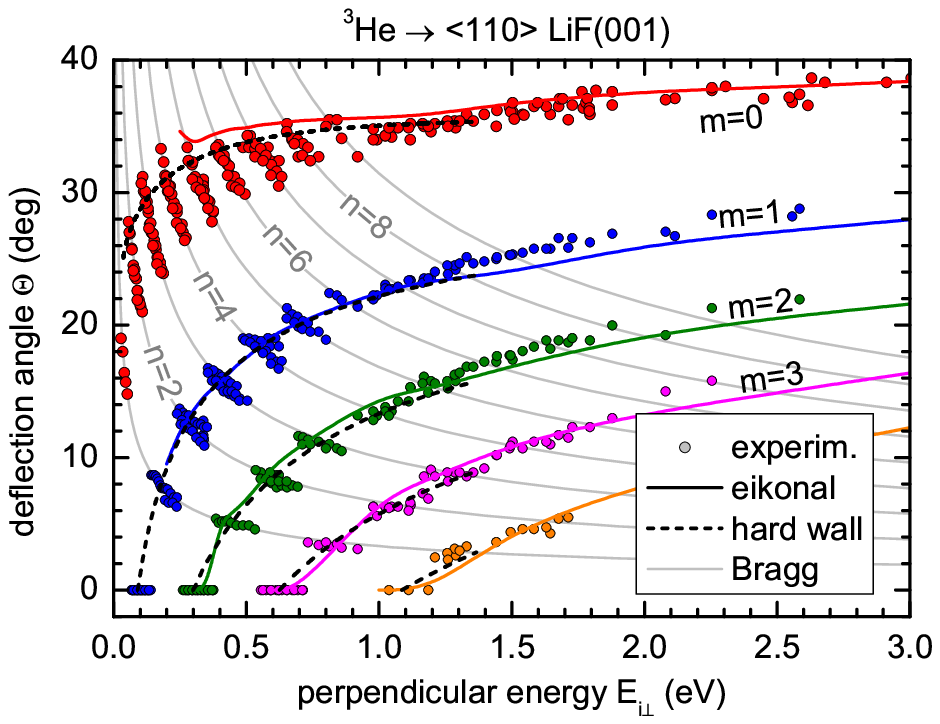}{\special{language
"Scientific Word";type "GRAPHIC";maintain-aspect-ratio TRUE;display
"USEDEF";valid_file "F";width 247.8125pt;height 198.25pt;depth
0pt;original-width 4.3033in;original-height 3.3252in;cropleft
"0.0322";croptop "0.9583";cropright "0.9194";cropbottom "0.0416";filename
'fig4.eps';file-properties "XNPEU";}}

Signatures of both interference processes can be observed in the simulated
spectrum \ also. In Fig. \ref{Fig5} the eikonal probability $dP/d\Theta $ is
plotted as a function of the deflection angle for $E_{i\perp }$ = 0.5 eV.
From Eq. (\ref{Teikn}) when the initial projectile position $\vec{R}_{os}$
is integrated over a unit cell, supernumerary maxima are only present in\
the angular projectile distribution. But when the integration area is
extended to include the first order nearest neighbor target ions, the
eikonal spectrum displays internal structures in the supernumerary maxima,
which are due to resolved Bragg peaks.

\FRAME{ftbpFU}{223.3125pt}{193.75pt}{0pt}{\Qcb{Eikonal results for $^{3}$He
atoms scattered from LiF(001) along the direction $\langle 110\rangle $ for
the perpendicular energy $E_{i\perp }$= 0.5 eV. Dashed (red) line, eikonal
differential probability derived by integrating the initial position over an
unit cell; solid (blue) line, similar by using an extended integration area,
as explained in the text. Dotted vertical lines, theoretical peak positions
based on the Bragg condition (Eq. \protect\ref{bragg}).}}{\Qlb{Fig5}}{%
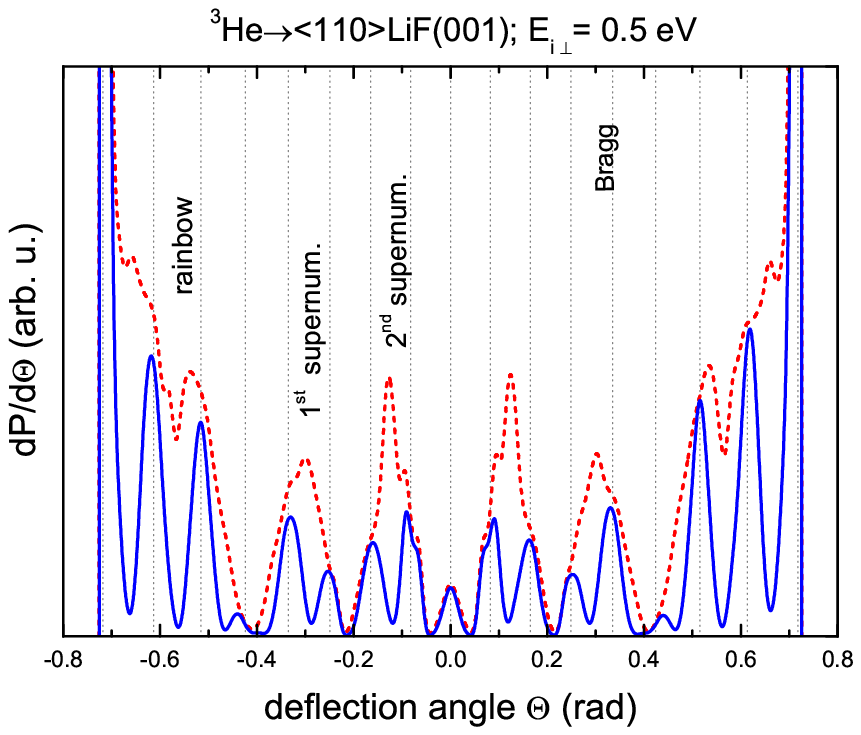}{\special{language "Scientific Word";type
"GRAPHIC";maintain-aspect-ratio TRUE;display "USEDEF";valid_file "F";width
223.3125pt;height 193.75pt;depth 0pt;original-width 4.0248in;original-height
3.3529in;cropleft "0.0516";croptop "0.9586";cropright "0.9138";cropbottom
"0.0620";filename 'fig5.eps';file-properties "XNPEU";}}

Since we are interested in studying the modulation of the potential inside
an unit cell, we have plotted in Fig. \ref{Fig4a} eikonal curves
corresponding to the center of supernumerary maxima, neglecting the Bragg
interference that appears as a superimposed structure at low perpendicular
energies. The eikonal curves obtained by using the DFT atom-surface
interaction follow closely the experimental results for supernumeraries $m=1$
to $m=4$ with only slight deviations for the quantum rainbow $m=0$ at low $%
E_{i\perp }$. Supernumerary rainbows are sensitive to the corrugation of the
equipotential surfaces but not to their positions. Each potential with the
same effective corrugation along the $\langle 110\rangle $ direction would
agree as well. In order to show that the potential is unique, one has to
compare the theoretical results with the experiment for a further
(different) channeling direction. The potential would be unique only if the
same potential describes the supernumeraries here as well.

In Fig. \ref{Fig4b} we show a comparison of experimental supernumerary
rainbows with eikonal results for scattering of $^{3}$He along a $\langle
100\rangle $ direction of LiF(001). The eikonal curves with DFT potential
agree with the experimental values except for $m=0$ at low energies $%
E_{i\perp }\lesssim $ 1.0 eV. We point out again that the eikonal
approximation is a semi-classical theory which fails in the vicinity of the
classical rainbow $\Theta _{rb}$. The angular position of the quantum
rainbow $m=0$ is smaller than the classical rainbow angle $\Theta _{rb}$
when $\lambda _{\perp }$ becomes large \cite{Garibaldi1975}. Since the
intensity at $\Theta _{rb}$ is overestimated the peak maximum $m=0$ is
shifted to larger deflection angles at lower $E_{i\perp }$. For a correct
description of the intensity at $\Theta _{rb}$ a \textquotedblleft uniform
approximation" \cite{Berry1966,Berry1972} is necessary, but supernumeraries
are not affected by this deviation. Since the positions of the
supernumeraries agree well in both channeling directions and since a change
of the corrugation of the He-LiF(001) interaction potential by 0.02 \AA\ %
induces a clear shift in the position of the supernumeraries \cite%
{Schuller08,Gravielle09}, we conclude that the DFT He-LiF(001) potential is
accurate.

\FRAME{ftbpFU}{248.0625pt}{199.875pt}{0pt}{\Qcb{ Similar to Fig. \protect\ref%
{Fig4a}, but for scattering along the $\langle 100\rangle $ direction.}}{%
\Qlb{Fig4b}}{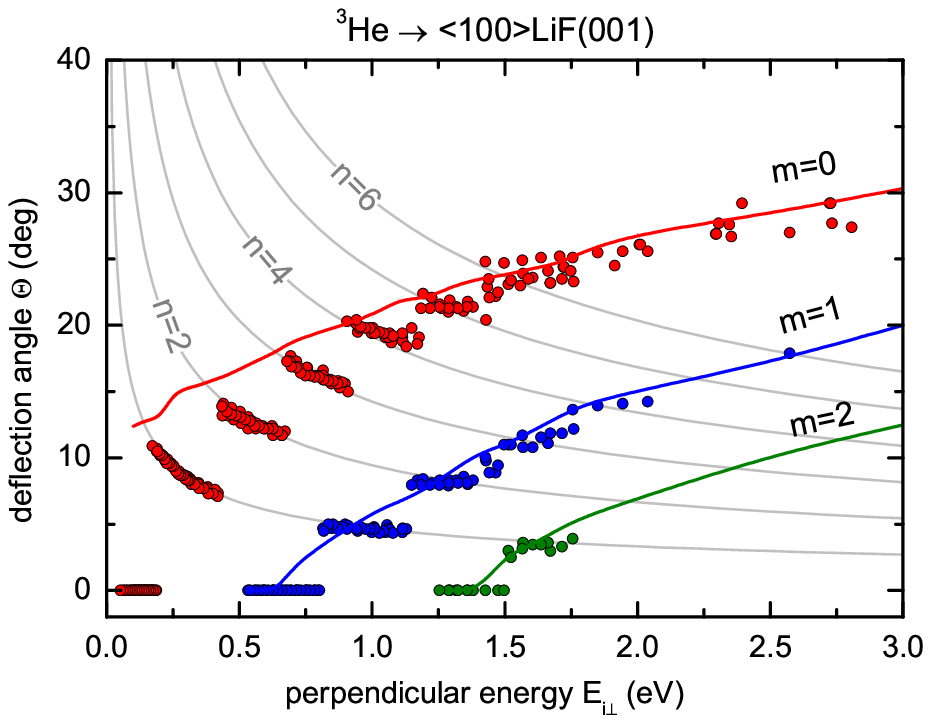}{\special{language "Scientific Word";type
"GRAPHIC";maintain-aspect-ratio TRUE;display "USEDEF";valid_file "F";width
248.0625pt;height 199.875pt;depth 0pt;original-width
4.2756in;original-height 3.3252in;cropleft "0.0324";croptop
"0.9583";cropright "0.9189";cropbottom "0.0416";filename
'fig6.eps';file-properties "XNPEU";}}

By employing the hard-wall model from Garibaldi \textit{et al.} \cite%
{Garibaldi1975} valid for sinusoidal small corrugated potential surfaces we
obtain good agreement as well. Hard wall approximations were applied
successful for description of scattering of He from LiF(001) with thermal
energies \cite{Boato73,Chow1976,Garcia1976a,Garcia1977b}. Due to the large
parallel velocity of the He projectile the effective potential is averaged
along the chains of atoms in the beam direction. The potential contours of
the effective potential for scattering along $\langle 110\rangle $ for low
energies are almost indistinguishable from sine functions. For a sinusoidal
hard wall, the intensity $I_{n}$ of a Bragg peak of order $n$ is given by: 
\begin{equation}
I_{n}=J_{n}^{2}\!\left( \frac{\pi \Delta z}{\lambda _{\mathrm{\perp }}}\!%
\left[ 1+\cos \Theta _{n}\right] \right) ,  \label{Bessel}
\end{equation}%
with $J_{n}$ being the Bessel function of order $n$, $\Theta
_{n}\!=\!\arccos \!\sqrt{1-(n\lambda _{\mathrm{\perp }}/d)^{2}}$ the
deflection angle of order $n$, and $\Delta z$ the full corrugation of the
sinusoidal hard wall, i.e. the normal distance between the maximum and the
minimum of a equipotential surface. In Ref. \onlinecite{Masel1976} it was
shown that the solution given by Eq. (\ref{Bessel}) is in agreement with the
exact quantum mechanical solution in a wide range of $\lambda _{\perp }$.
Since we are interested in the angular positions of the supernumerary
rainbows we treat $n$ in Eq. (\ref{Bessel}) as $\in \mathbb{R}$ and search
for the maxima of this oscillating function by searching the zeros of the
derivation of Eq. (\ref{Bessel}) where for $\Delta z(E_{i\perp })$ the
effective corrugation of the present DFT potential across the $\langle
110\rangle $ channel was used. Results are displayed in Fig. \ref{Fig4a},
showing good agreement with experiment and with eikonal results. It allows
us to conclude that the hard wall model is a good approximation here.

\FRAME{ftbpFU}{207.9375pt}{264.4375pt}{0pt}{\Qcb{Angular distributions for
intensities projected inside the annulus (cf. Fig.1) for scattering of He
atoms from LiF(001) along \TEXTsymbol{<}110\TEXTsymbol{>} under $\protect%
\theta _{i}$ = 0.99 deg. Upper panel: $E_{i}$ = 0.35 keV, middle panel: $%
E_{i}$ = 0.50 keV, lower panel: $E_{i}$ = 0.65 keV. Solid curves represent
best fits to data by sum of peaks with Lorentzian lineshapes. Numbers denote
diffraction orders. }}{\Qlb{Fig8}}{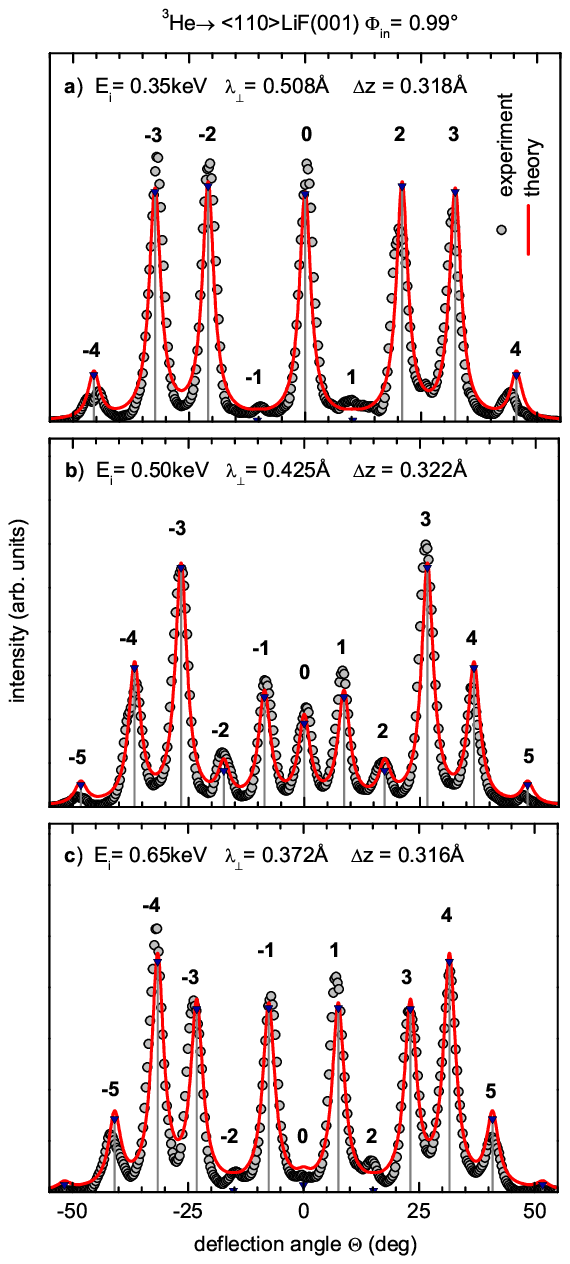}{\special{language "Scientific
Word";type "GRAPHIC";display "USEDEF";valid_file "F";width 207.9375pt;height
264.4375pt;depth 0pt;original-width 3.2508in;original-height
5.5279in;cropleft "0.1212";croptop "0.9642";cropright "0.8484";cropbottom
"0.0355";filename 'fig7.eps';file-properties "XNPEU";}}

We then apply the hard wall approximation in order to describe the different
relative intensities of the Bragg peaks as shown in Fig. \ref{Fig8}. We fit
a sum of Lorentzian peaks to the diffraction pattern, where the peak
positions are given by the Bragg relation (Eq. \ref{bragg}). We assume, that
the peak width is the same for all diffraction spots \cite{Busch2009}.
Resulting relative peak heights are compared with the relative intensities
given by Eq. (\ref{Bessel}) with $\Delta z$ as a fit parameter. The
resulting intensity distributions are shown as red/gray curves in Fig. \ref%
{Fig8}. The values for $\Delta z$ obtained from the best fit are given in
each panel. Since $\Delta z$ is almost constant, the differences in the
intensities are due to the different de Broglie wavelength $\lambda _{\perp
} $ only.

\FRAME{ftbpFU}{229.0625pt}{184.9375pt}{0pt}{\Qcb{ Effective corrugation $%
\Delta z$ of the potential across the $\langle 110\rangle $ channel as a
function of the perpendicular energy $E_{i\perp }$. Open symbols,
experimentally-derived results in the hard wall approximation; solid red
curve, values derived from the present DFT potential; dashed curve,
corrugation obtained from the interaction potential from Celli \textit{et al.%
} \protect\cite{Celli85}; dashed dotted curve, corrugation from the
potential from Celli \textit{et al.} but with the correct He-F$^{-}$pair
potential from Erratum of Ahlrichs \textit{et al.} \protect\cite%
{Ahlrichs1993}.}}{\Qlb{Fig6}}{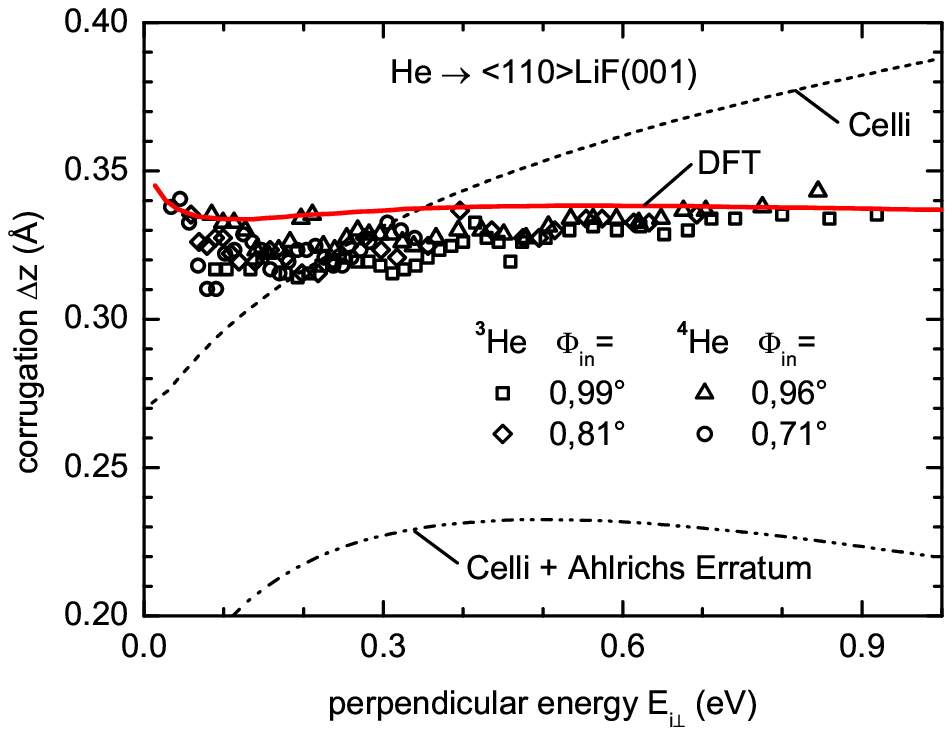}{\special{language "Scientific
Word";type "GRAPHIC";maintain-aspect-ratio TRUE;display "USEDEF";valid_file
"F";width 229.0625pt;height 184.9375pt;depth 0pt;original-width
4.4555in;original-height 3.5613in;cropleft "0.0465";croptop
"0.9221";cropright "0.9221";cropbottom "0.0388";filename
'fig8.eps';file-properties "XNPEU";}}

The effective corrugation $\Delta z$ of the potential across the $\langle
110\rangle $ channel deduced from Bragg peak intensities for scattering of $%
^{3}$He and $^{4}$He under different angles of incidence is plotted as a
function of the perpendicular energy $E_{i\perp }$ in Fig. \ref{Fig6} as
open symbols. Each data point corresponds to a best fit with Eq. (\ref%
{Bessel}) as in Fig. \ref{Fig8}. Although the diffraction patterns are
different for the He isotopes and different angles of incidence, the data
are well described by Eq. (\ref{Bessel}) with an almost constant corrugation 
$\Delta z$. We compare the experimentally derived values with the effective
corrugation of the present DFT potential (red/gray full curve in Fig. \ref%
{Fig6}) defined as the normal distance between the maximum and minimum of
the equipotential surface obtained by averaging the surface potential along
the $\langle 110\rangle $ axial channel. We observe that the theoretical
curve is close to the experimental data.

Currently, the interaction potential from Celli \textit{et al.} \cite%
{Celli85} is considered as the best available He-LiF(001) potential \cite%
{Jardine2004,Riley2007}. This potential is actually constructed for
interactions at thermal energies but the analytical expression can be
evaluated also for energies of some eV. The resulting effective corrugation
shows a strong dependence on the perpendicular energy $E_{i\perp }$ in
contrast to the experimental values. We conclude that the potential from
Celli \textit{et al.} is not adequate for the description of the He-LiF
interaction in the eV range. This is consistent with the results of Ref. 
\cite{Jardine2004} and \cite{Riley2007} where the experiment is sensitive to
the attractive potential well of the planar averaged potential in the meV
range only, but not to the corrugation of the repulsive part of the
interaction potential in the eV domain.

The repulsive part of the Celli potential is based on SCF pair potentials
for He-Li$^{+}$ and He-F$^{-}$ from Ahlrichs \textit{et al.} \cite%
{Ahlrichs1988}. Due to an error in the calculation of He-F$^{-}$ repulsion 
\cite{Ahlrichs1993}, which is the dominant contribution of the repulsive
part of the He-LiF(001) interaction potential, the Celli potential has to be
corrected. The correct He-F$^{-}$ potential \cite{Ahlrichs1988,Ahlrichs1993}
is in good agreement with recent He-F$^{-}$ \textit{ab-initio} pair
potentials \cite{Archibong1998,Gray2006} as well as the original He-Li$^{+}$
pair potential \cite{Ahlrichs1988} agrees with recent He-Li$^{+}$ \textit{%
ab-initio} potentials \cite{Soldan2001}. Nevertheless, inserting the
corrected He-F$^{-}$ parameters in the expression for the Celli potential,
the effective corrugation $\Delta z(E_{i\perp })$ deviates from the
experimental results even more.

\FRAME{ftbpFU}{225.3125pt}{193.1875pt}{0pt}{\Qcb{Angular distributions for
He atoms scattered from LiF(001) along the direction $\langle 110\rangle $
with $\protect\lambda _{\perp }=$ 0.13 \AA . Two different isotopes and
impact energies are considered. Red circles and full curve, experimental
data and eikonal results, respectively, for 3.5 keV $^{3}$He; blue squares
and dashed curve, for 2.6 keV $^{4}$He.}}{\Qlb{Fig7}}{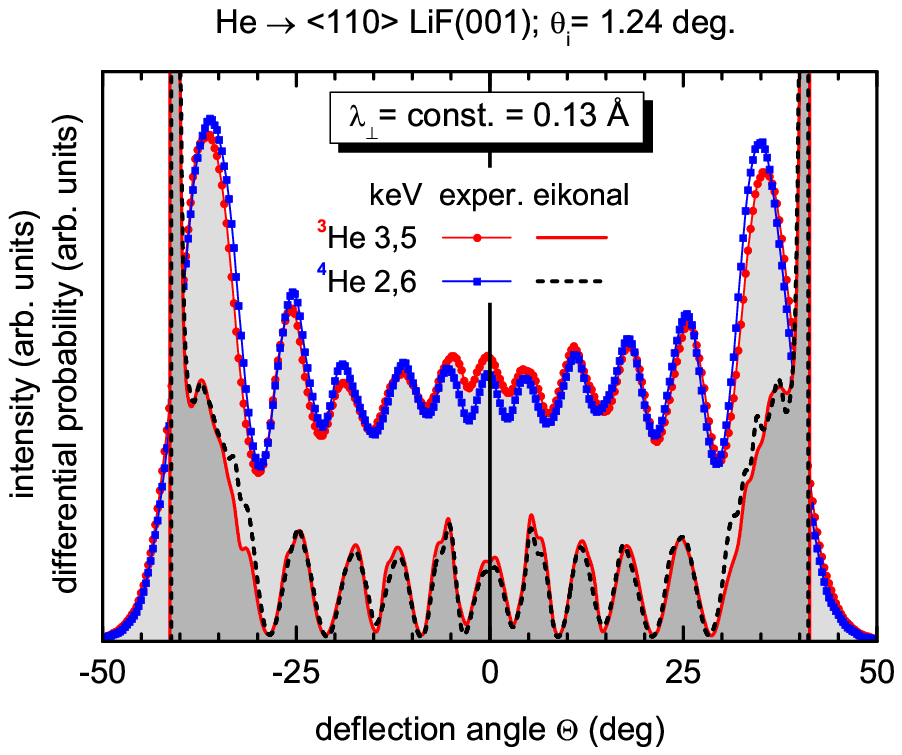}{\special%
{language "Scientific Word";type "GRAPHIC";maintain-aspect-ratio
TRUE;display "USEDEF";valid_file "F";width 225.3125pt;height
193.1875pt;depth 0pt;original-width 4.19in;original-height 3.4515in;cropleft
"0.0330";croptop "0.9598";cropright "0.9172";cropbottom "0.0401";filename
'fig9.eps';file-properties "XNPEU";}}

The small variation of the effective corrugation with the perpendicular
energy explains why He isotopes with the same $\lambda _{\perp }$ but
different perpendicular energies $E_{i\perp }$ produce similar diffraction
patterns, even though different regions of the surface potential are probed.
As observed in Fig. \ref{Fig7}, helium atoms with the same perpendicular de
Broglie wavelength ($\lambda _{\perp }=$ 0.13 \AA ) but different
perpendicular energies ($E_{i\perp }$=1.63 eV and 1.21 eV) show identical
eikonal distributions, as in the experiment. The scattering processes take
place indeed at different distances to the surface, but since $\lambda
_{\perp }$ is the same, similar interference patterns appear.

\section{Conclusions}

We have studied diffraction patterns for swift He atoms colliding grazingly
with a LiF(001) surface in order to test the \textit{ab-initio} surface
potential, obtained from DFT by making use of the \textit{Siesta} code.
Angular spectra of scattered projectiles are obtained with the DFT potential
from the surface eikonal approximation, which takes into account the quantum
interference due to the coherent superposition of transition amplitudes
corresponding to different projectile paths with the same deflection angle.

For incidence along the $\langle 110\rangle $ channel, the dependence on the
parallel and perpendicular components of the impact energy was analyzed. It
was found that angular distributions in terms of the deflection angle $%
\Theta $ are completely governed by the perpendicular energy. Diffraction
spectra as a function of $\Theta $ give information about the surface
potential across the incidence channel by means of two different mechanisms,
supernumerary rainbow and Bragg interferences. We have focused here on
supernumerary rainbows which are very sensitive to the corrugation of the
surface potential within a unit cell. By comparison of eikonal angular
spectra with experimental distributions we concluded that the DFT model
provides a good description of the surface potential for perpendicular
energies in the range from $E_{i\perp }$=0,03 eV up to 3 eV. The agreement
between theoretical and experimental results for the intensity near the
classical rainbow angle is poorer for smaller $E_{i\perp }$. This deficiency
is attributed to the range of validity of the semiclassical models, like the
eikonal approach, which do not include quantum effects with respect to the
projectile trajectory.

For scattering along the $\langle 110\rangle $ direction the calculation in
the hard wall approximation using Eq. (\ref{Bessel}) is in good agreement
with results from simulation with the soft potential based on the eikonal
model. We conclude that in this special case the hard wall approximation is
good.

We found that the potential from Celli \textit{et al.} \cite{Celli85} is not
adequate for the description of the interaction of He with LiF(001) in the
eV energy domain. The DFT potential displays an almost constant effective
corrugation $\Delta z$ as a function of the perpendicular energy, in
agreement with the experimental values. As a consequence of this nearly
constant corrugation, He isotopes with the same perpendicular de Broglie
wavelength but consequently different perpendicular energies give rise to
the same interference patterns.

\begin{acknowledgments}
M.S.G and J.E.M acknowledge financial support from CONICET, UBA, and ANPCyT
of Argentina. JMP acknowledges financial support from Spain's Ministry of
Education and CSIC under the JAE-Doc program. A.S. thanks the IMPRS-CS PhD
program of the MPG for financial support. The assistance of S. Wethekam and
K. Maass in running the experiments is gratefully acknowledged. This work is
supported by the Deutsche Forschungsgemeinschaft under DFG contract No. Wi
1336.
\end{acknowledgments}


\begin{thebibliography}{99}
\bibitem{Cabrera70} N. Cabrera, V. Celli, F.O. Goodman, and R. Manson, Surf.
Sci. \textbf{19,} 67 (1970).

\bibitem{Boato73} G. Boato \textit{et al}., J. Phys. C \textbf{6,} L394
(1973).

\bibitem{Boato1976a} G. Boato, P. Cantini, L. Mattera, Surf. Sci. \textbf{55,%
} 141 (1976).

\bibitem{Wolken1973} G. Wolken, J. Chem. Phys. \textbf{58} 3047 (1973).

\bibitem{Garcia1976b} N. Garc\'{\i}a, J. Ib\'{a}ez, J. Solana, N. Cabrera,
Solid State Communications \textbf{20}, 1159 (1976).

\bibitem{Garcia1977b} N. Garcia, J. Chem. Phys. \textbf{67,} 897 (1977).

\bibitem{Hubbard83} L. M. Hubbard and W.H. Miller, J. Chem. Phys. \textbf{78}
1801 (1983).

\bibitem{Celli85} V. Celli, D. Eichenauer, A. Kaufhold, and J.P. Toennies,
J. Chem. Phys. \textbf{83}, 2504 (1985).

\bibitem{Ekinci2004} Y. Ekinci, J. P. and Toennies, Surf. Sci. \textbf{563},
127 (2004).

\bibitem{Jardine2004} A. P. Jardine \textit{et al}., Science \textbf{304}
1790 (2004).

\bibitem{Riley2007} D. J. Riley, A. P. Jardine, S. Dworski, G.
Alexandrowicz, P. Fouquet, J. Ellis, W. Allison, J. Chem. Phys. \textbf{126}%
, 104702 (2007).

\bibitem{Schuller07} A. Sch\"{u}ller, S. Wethekam, and H. Winter, Phys. Rev.
Lett. \textbf{98,} 016103 (2007).

\bibitem{Rousseau07} P. Rousseau, H. Khemliche, A.G. Borisov, and P. Roncin,
Phys. Rev. Lett. \textbf{98} 016104 (2007).

\bibitem{Schuller08} A. Sch\"{u}ller and H. Winter, Phys. Rev. Lett. \textbf{%
100,} 097602 (2008).

\bibitem{Schuller2009} A. Sch\"{u}ller, M. Busch, S. Wethekam, H. Winter,
Phys. Rev. Lett. \textbf{102} 017602 (2009).

\bibitem{Gravielle08} M.S. Gravielle and J.E. Miraglia, Phys. Rev. A. 
\textbf{78}, 022901 (2008).

\bibitem{Aigner08} F. Aigner, N. Simonovi\'{c}, B. Solleder, L. Wirtz, and
J. Burgd\"{o}rfer, Phys. Rev. Lett. \textbf{101,} 253201(2008).

\bibitem{Schuller2009PRB} A. Sch\"{u}ller, M. Busch, J. Seifert, S.
Wethekam, H. Winter, K. G\"{a}rtner, Phys. Rev. B \textbf{79,} 235425 (2009).

\bibitem{Schuller09} A. Sch\"{u}ller and H. Winter, Nucl. Instrum. Meth.
Phys. Res. B \textbf{267}, 628 (2009).

\bibitem{siesta} J.M. Soler, E. Artacho, J.D. Gale, A. Garc\'{\i}a, J.
Junquera, P. Ordej\'{o}n and D. S\'{a}nchez-Portal, J. Phys.: Condens.
Matter \textbf{14}, 2745-2779 (2002).

\bibitem{siesta1} E. Artacho, E. Anglada, O. Dieguez, J. D. Gale, A. Garc%
\'{\i}a, J. Junquera, R. M. Martin, P. Ordej\'{o}n, J. M. Pruneda, D. S\'{a}%
nchez-Portal and J. M. Soler, J. Phys.: Condens. Matter 20, 064208 (2008).

\bibitem{Gravielle09} M.S. Gravielle and J.E. Miraglia, Nucl. Instrum. Meth.
Phys. Res. B \textbf{267}, 610 (2009).

\bibitem{CA} J. P. Perdew and A. Zunger, Phys. Rev. B \textbf{23}, 5048
(1981).

\bibitem{TM} N. Troullier and J. L. Martins, Phys. Rev. B \textbf{46}, 1754
(1992).

\bibitem{Joachain} C.J. Joachain, \textit{Quantum Collision Theory}
(North-Holland, Amsterdam, 1979).

\bibitem{Avrin94} W.F. Avrin and R.P. Merrill, Surf. Sci. \textbf{311,} 269
(1994).

\bibitem{McDowellColeman} M.R.C. McDowell and J.P. Coleman, \textit{%
Introduction to the Theory of Ion-Atom Collisions} (North-Holland,
Amsterdam, 1970).

\bibitem{Bederson} T. M. Miller and B. Bederson, \textit{Advances in Atomic
and Molecular Physics}, Vol. \textbf{13}, 1-55, edited by D. R. Bates and B.
Bederson (Academic, New York, 1977).

\bibitem{Vogt02} J.Vogt and H. Weiss, Surf. Sci. \textbf{501,} 203 (2002).

\bibitem{BSSE} F. Boys and F. Bernardi, Mol. Phys. \textbf{19}, 553 (1970).

\bibitem{Berry1972} M. V. Berry and K. E. Mount, Reports on Progress in
Physics \textbf{35,} 315 (1972).

\bibitem{Berry1966} M. V. Berry, Proceedings of the Physical Society \textbf{%
89,} 479 (1966)

\bibitem{Garibaldi1975} U. Garibaldi, A. C. Levi, R. Spadacini, G. E.
Tommei, Surf. Sci. \textbf{48,} 649 (1975).

\bibitem{Masel1976} R. I. Masel, R. P. Merrill, W. H. Miller, J. Chem. Phys. 
\textbf{65,} 2690 (1976).

\bibitem{Chow1976} H. Chow, E. D. Thompson, Surf. Sci. \textbf{54,} 269
(1976).

\bibitem{Garcia1976a} N. Garc\'{\i}a, J. Ib\'{a}\~{n}ez, J. Solana, N.
Cabrera, Surf. Sci. \textbf{60,} 385 (1976).

\bibitem{Busch2009} M. Busch, A. Sch\"{u}ller, S. Wethekam, H. Winter, Surf.
Sci. \textbf{603,} L23 (2009).

\bibitem{Ahlrichs1988} R. Ahlrichs, H. J. Bohm, S. Brode, K. T. Tang, J. P.
Toennies, J. Chem. Phys. \textbf{88,} 6290 (1988).

\bibitem{Ahlrichs1993} R. Ahlrichs, H. J. Bohm, S. Brode, K. T. Tang, J. P.
Toennies, J. Chem. Phys. \textbf{98,} 3579 (1993).

\bibitem{Archibong1998} E. F. Archibong, C. Hu, A. J. Thakkar, J. Chem.
Phys. \textbf{109,} 3072 (1998).

\bibitem{Gray2006} B. R. Gray, T. G. Wright, E. L. Wood, L. A. Viehland,
Phys. Chem. Chem. Phys. \textbf{8,} 4752 (2006).

\bibitem{Soldan2001} P. Sold\'{a}n, E. P. F. Lee, J. Lozeille, J. N.
Murrell, T. G. Wright, Chemical Physics Letters \textbf{343,} 429 (2001).
\end{thebibliography}
\end{document}